\documentclass{optica-article}

\journal{oe}

\articletype{Research Article}

\usepackage{multirow}

\usepackage{enumitem}

\usepackage{siunitx}

\usepackage{placeins}

\usepackage{listings}
\usepackage{color}  

\definecolor{codegreen}{rgb}{0,0.6,0}
\definecolor{codegray}{rgb}{0.5,0.5,0.5}
\definecolor{codepurple}{rgb}{0.58,0,0.82}
\definecolor{backcolour}{rgb}{0.85,0.85,0.82}

\lstdefinestyle{mystyle}{
	backgroundcolor=\color{backcolour},   
	commentstyle=\color{codegreen},
	keywordstyle=\color{magenta},
	numberstyle=\tiny\color{codegray},
	stringstyle=\color{codepurple},
	basicstyle=\ttfamily\footnotesize,
	breakatwhitespace=false,
	breaklines=false,
	captionpos=b,
	keepspaces=true,
	numbers=left,
	numbersep=5pt,
	showspaces=false,
	showstringspaces=false,
	showtabs=false,
	tabsize=2
}
\usepackage{chngcntr}

\newcommand*{\br}{\ensuremath{\mathbf{r}}}
\newcommand*{\bw}{\ensuremath{\mathbf{w}}}
\newcommand*{\bu}{\ensuremath{\mathbf{u}}}
\newcommand*{\sm}{\ensuremath{\mathbf{S}_m}}

\begin{document}

\title{High-speed processing of X-ray wavefront marking data with the Unified Modulated Pattern Analysis (UMPA) model}

\author{
	Fabio~De~Marco,\authormark{1,2,*}
	Sara~Savatovi\'{c},\authormark{1,2}
	Ronan~Smith,\authormark{3}
	Vittorio~Di~Trapani,\authormark{1,2}
	Marco~Margini,\authormark{1,2}
	Ginevra~Lautizi,\authormark{1,2}
	and Pierre~Thibault\authormark{1,2}}

\address{\authormark{1}Department of Physics, University of Trieste, Via~Valerio~2, 
34127~Trieste, Italy\\
\authormark{2}Elettra-Sincrotrone Trieste, Strada Statale 14 -- km 163.5, 34149~Basovizza, Italy\\
\authormark{3}Department of Physics, University of Southampton, University Road, Southampton, SO17~1BJ, UK}

\email{\authormark{*}fabiodomenico.demarco@units.it} 



\begin{abstract}
Wavefront-marking X-ray imaging techniques use e.g., sandpaper or a grating to generate intensity fluctuations, and analyze their distortion by the sample in order to retrieve attenuation, phase-contrast, and dark-field information.
Phase contrast yields an improved visibility of soft-tissue specimens, while dark-field reveals small-angle scatter from sub-resolution structures.
Both have found many biomedical and engineering applications.
The previously developed Unified Modulated Pattern Analysis (UMPA) model extracts these modalities from wavefront-marking data.
We here present a new UMPA implementation, capable of rapidly processing large datasets and featuring capabilities to greatly extend the field of view. We also discuss possible artifacts and additional new features.
\end{abstract}


\section{Introduction}
In conventional X-ray imaging, information retrieved from a sample is encoded purely by its 
attenuation of the incident light.
X-ray phase-contrast imaging, on the other hand, retrieves information of an imaged sample by 
examining its effect on the phase of an incident wavefront.
Since the attenuation of very thin samples, or samples composed of light elements, is often weak, 
resulting attenuation images can exhibit very low contrast.
In such cases, X-ray phase-contrast techniques often provide images with a superior signal-to-noise 
ratio, since the real decrement $\delta$ of the index of refraction exceeds its imaginary part $\beta$ 
by several orders of magnitude~\cite{Momose2005}.
As X-ray detectors cannot determine the phase of a wavefront, X-ray phase-contrast imaging 
methods encode this information in an intensity pattern.

Over the last three decades, a wide range of methods for achieving such an encoding have been 
developed.
Among these are ``propagation-based imaging'', which exploits self-interference of the beam after 
being disturbed by a sample~\cite{Snigirev1995,Cloetens1999,Paganin2002,Gureyev2020}, 
``analyzer-based imaging'', which isolates wave components with a specific propagation direction 
through the use of an analyzer crystal~\cite{Chapman1997,Oltulu2003}, ``grating-based imaging'', 
where the distortion of Talbot self-images produced by a transmission grating is 
analyzed~\cite{Momose2003,Pfeiffer2006,Pfeiffer2008}, the closely-related, but 
non-interferometric ``edge-illumination'' technique~\cite{Olivo2001,Munro2012,Endrizzi2014}, and 
finally a range of ``wavefront-marking'' methods, where a single optical element generates a 
directly resolved intensity modulation, whose distortion is then algorithmically analyzed.
Besides attenuation and phase information, several of these 
methods~\cite{Oltulu2003,Pfeiffer2008,Endrizzi2014,Gureyev2020} also retrieve ``dark-field'' 
information, which characterizes the amount of small-angle scattering.

A wide range of optical elements may be used for wavefront marking: besides ordered, periodic 
objects such as gratings~\cite{Morgan2010, Gustschin2021}, randomly-oriented structures such as 
sandpaper may also be used~\cite{Berujon2012a,Berujon2012b,Morgan2012}.
The latter approach produces (near-field) X-ray speckle~\cite{Cerbino2008}, hence the method is 
also called ``speckle-based imaging''.
An important benefit of the method is its experimental simplicity, and its ability to retrieve both 
differential-phase and dark-field information in two dimensions.
Despite its relatively recent introduction, it has found applications in wavefront sensing, 
at-wavelength metrology of X-ray optics components, as well as biomedical and materials science 
imaging problems~\cite{Zdora2018, Berujon2020b}.

We here give a brief overview of existing approaches for extracting information from X-ray speckle 
imaging data.
Two novel approaches have recently been demonstrated for extracting information from 
speckle-based X-ray data: firstly, the "ptychographic X-ray speckle tracking" (PXST) technique which 
iteratively reconstructs sample and wavefront from a series of images where the sample is laterally translated \cite{Morgan2020b,Morgan2020a}.
Secondly, from the insight that the Fokker-Planck differential equation can be used to model refraction and dark-field~\cite{Paganin2019, Morgan2019}, ``multi-modal intrinsic speckle-tracking'' (MIST) has been developed. The distortion of speckle patterns is quantified by solving this equation~\cite{Pavlov2020}, even allowing the extraction of directional scatter~\cite{Pavlov2021}.

Earlier processing approaches instead attempt to correlate subregions of sample and reference 
images.
Two reviews give a more in-depth insight to these different approaches~\cite{Zdora2018, 
Berujon2020a}.
They are derived from the differential interference contrast (DIC) technique, which uses laser 
speckle to characterize strain in a material~\cite{Pan2009}.

In ``X-ray speckle tracking'' (XST), one image each is taken with and without the sample.
To identify the displacement in a pixel of the sample image, a neighborhood of this pixel, the 
``(analysis) window'', is identified and the best match of this window in the reference image is sought 
by maximizing the zero-normalized cross-correlation (ZNCC) between the windows.
The 2D displacement vector between these windows represents the local amount of refraction.
Since the analysis window must be greater than one pixel, its extent limits the spatial resolution of 
resulting images ~\cite{Berujon2012a, Morgan2012}.

In ``X-ray speckle scanning'' (XSS), the diffuser is displaced on a very fine 1D or 2D grid, yielding a 
1D or 2D map of intensities for each pixel (with and without the sample).
The displacement is calculated by identifying the shift between the two intensity maps.
A higher spatial resolution and better angular phase sensitivity can be achieved, albeit at the 
expense of a slower acquisition and the need for very precise diffuser 
positioning\cite{Berujon2012b}.

In ``X-ray speckle vector tracking'' (XSVT), the diffuser is displaced in intervals of arbitrary shape, 
which yields a vector of intensities for each detector pixel (one value per diffuser position).
For each sample image vector, a reference image vector is sought that maximizes the ZNCC.
The shift is then given as the distance between the vectors' pixel locations.
Attenuation and dark-field data are also retrieved~\cite{Berujon2015}. The method was later 
extended to also use an analysis window~\cite{Berujon2016}.

An alternative to maximizing the ZNCC is minimizing a cost function which, in addition to lateral 
shift, also incorporates attenuation and dark-field information.
This approach was first applied in~\cite{Zanette2014}, and later generalized to the ``Unified 
Modulated Pattern Analysis'' (UMPA) model by the inclusion of diffuser stepping with arbitrary 
trajectories (like XSVT)~\cite{Zdora2017}.

An important limitation of most of the above-mentioned methods is the significant required 
computational effort.
This is especially severe in the case of tomography, which requires a large number of projections.
The problem is compounded by the steady increase in detector resolution and frame rate, and thus 
the generated data volume, especially at synchrotron beamlines.

In order to make the analysis of speckle imaging datasets more accessible, we here present an 
improved implementation of the UMPA model.
Compared to the previous version~\cite{Zdora2017}, a number of new features have been introduced, and single-thread execution speed has been increased by two orders of magnitude.

Most importantly, the aspect of displacing the diffuser between measurements has been generalized to instead allow displacing the sample (similar to the technique presented in \cite{Morgan2020b}).
By defining a suitable sample ``trajectory'', the effective field of view can be arbitrarily increased, 
limited only by the motor travel range.
We demonstrate the package’s capabilities on sample-stepping measurements performed at two 
different synchrotron beamlines.

The speed increase is achieved by translation of the core routines to C++, while the ease of use of 
previous Python versions is maintained by embedding the C++ code in a Python interface through 
Cython~\cite{Behnel2011}.
Speed is further increased by parallelization with OpenMP.

We also discuss the presence of a bias in the estimation of retrieved speckle shifts and propose a 
simple method to reduce its magnitude.
Additionally, since matching sections of the speckle patterns in reference and sample image data 
sets are shifted relative to one another, the assignment of a found signal value to a pixel location is 
ambiguous. We discuss and compare the two obvious approaches.
Finally, we present newly introduced region-of-interest and masking features.

We introduce the UMPA model in section~\ref{sec:math}.
The minimization of the model's cost function (which has changed significantly compared to 
the previous version) is discussed in section~\ref{sec:minimization}.
The above-mentioned additions and improvements compared to the previous version are 
introduced with example measurements in section~\ref{sec:improvements}.
A brief conclusion and outlook are given in section~\ref{sec:conclusion}.
Further mathematical details and a code example are provided in the appendices.

\section{Mathematical model of UMPA}\label{sec:math}

UMPA is a template-matching algorithm that compares intensity-modulated patterns through the use 
of analysis windows. In addition to lateral displacements, local variations in mean intensity and 
modulation amplitude between the patterns are also detected.
The mathematical foundations have previously been described in \cite{Zdora2017}.
We re-introduce the key features of the model here, and also include extensions added by the new 
implementation.
All variables in boldface represent two-dimensional vectors in the image plane.

We assume that a single image $I_0(\br)$ [$\br=(r_x, r_y)$] is acquired with a diffuser, but without a sample in the beam (the ``reference image'').
After introducing the sample, another image $I(\br)$ (the ``sample image'') is acquired.
UMPA performs a comparison between $I_0$ and $I$ for every pixel $\br$, by centering an analysis window of $(2N+1)\times (2N+1)$ pixels ($N = 0, 1, 2, \ldots$) on $I(\br)$, and seeking the most similar match of this window in the reference image $I_0(\br)$.
In the absence of attenuation and dark-field, as well as negligible change of wavefront curvature due to the sample, the effect of the sample is purely described by a local transverse translation $\bu=(u_x, u_y)$ of the image content.
The effect of attenuation is described by a uniform decrease of intensity, whereas dark-field is 
modeled as a decrease in the modulation amplitude, which we define as the difference of the 
speckle pattern to its local mean.
In short, the effect of the sample is modeled as:
\begin{equation}\label{eq:UMPA_model}
	\begin{split}
		I^\text{(model)}(\br; \bu, T, D) &= T \left\{ D \left[ I_{0}(\br - \bu) - \langle I_{0} \rangle(\br - 
		\bu) \right]
		+ \langle I_{0} \rangle(\br - \bu) \right\}, \\
		\langle I_{0} \rangle(\br) &= \frac{\sum_{w_x=-N}^{N} \sum_{w_y=-N}^{N} \Gamma(\bw) 
		I_{0}(\br+\bw)}{\sum_{w_x=-N}^{N} \sum_{w_y=-N}^{N} \Gamma(\bw)}.
	\end{split}
\end{equation}

Here, $\bw = (w_x, w_y)$ parameterizes the summation over the analysis window. The window 
function $\Gamma$ (a 2D Hamming window) is used to gradually decrease the weight of pixels 
towards the edge of the window. A simpler model without dark-field, which does not 
require the calculation of $\langle I_{0} \rangle(\br)$, is easily obtained by setting $D=1$ in the 
above equation.
UMPA attempts to minimize the sum of squared differences (SSD) between the reference speckle 
images $I_0$, modulated according to Eq.~\eqref{eq:UMPA_model}, and the speckle images $I$ 
acquired with the sample. Furthermore, information from multiple images with different lateral 
displacements between sample and diffuser may be combined.
Thus, the estimated images $\mathbf{\hat{u}}$, $\hat{T}$, $\hat{D}$ are those for which
\begin{equation}\label{eq:cost_function_simple}
	\begin{split}
		L(\br; \bu, T, D) = \sum_{m=1}^{M} \sum_{w_x, w_y=-N}^{N} \Gamma(\bw) \left[ 
		I_m^\text{(model)}(\br+\bw-\mathbf{S}_m; \bu, T, D) - I_m(\br+\bw-\mathbf{S}_m) \right]^2
	\end{split}
\end{equation}
is minimized.
The subscript $m$ denotes the images with the $m$-th diffuser-to-sample position.
The quantity $\Gamma(\bw)$ [same as in the definition of $\langle I_{0} \rangle(\br)$] here serves to smooth the effect which sharp edges in the input images may have on the output signals.

In the sample-stepping technique (introduced and discussed in section~\ref{sec:diffuser_sample_stepping}), the sample is displaced between frames instead of the diffuser.
This lateral displacement $\mathbf{S}_m$ of the $m$-th frame must be reverted to ensure that, for each $m$, the same sample feature contributes to $L$ in any given position $\br$.
For the conventional diffuser-stepping approach, $\mathbf{S}_m = \mathbf{0}$ for all $m$.
The values found in the minimization (see section~\ref{sec:minimization}) are then assigned to the position $\br$:
\begin{equation}
	\begin{split}
		\mathbf{\hat{u}}(\br), \hat{T}(\br), \hat{D}(\br) = \ &\underset{\bu, T, 
		D}{\operatorname{argmin}} \ L(\br; \bu, T, D).
	\end{split}
\end{equation}

\section{Minimization of the cost function}\label{sec:minimization}
The cost function of the model introduced above has four variables (three if $D$ is fixed to $1$), 
which suggests that its minimization produces a trajectory in a four-dimensional (or three-dimensional) parameter space.
In the form presented above however, the introduced models are only well-defined for integer values of $u_x$ and $u_y$, whereas $T$ and $D$ are floating-point numbers.
Most common optimization algorithms are not suitable for parameter spaces of such a structure.

To solve this problem, we introduce a partial optimization procedure, namely the minimization of the cost function with respect to $T$ and $D$ (or only $T$ if $D=1$), for a fixed shift $\bu$.
Due to the mathematical structure of the UMPA cost function, this step can be performed 
analytically (calculation details are provided in Appendix~\ref{sec:appendix_partial_opt}.
By performing this step for each estimate of $\bu$, optimal values for $T$ and $D$ ($\hat{T},\ \hat{D}$), and a cost function value minimized with respect to $T$ and $D$ [$\hat{L}(\br;\bu) = L(\br; \bu, \hat{T}, \hat{D})$] are calculated.
The problem is thus reduced to an optimization in $\mathbb{Z}^2$ ($\bu$-space) for each pixel.

In order to determine the sample shift $\hat{\bu}(\br)$, the cost function's global minimum, i.e.
\begin{equation}\label{eq:u_minimize}
	\begin{split}
		\hat{\bu}(\br) = \underset{\bu}{\operatorname{argmin}} \ \hat{L}^{(T,D)}(\br; \bu)
		\qquad \text{or} \qquad
		\hat{\bu}(\br) = \underset{\bu}{\operatorname{argmin}} \ \hat{L}^{(T)}(\br; \bu)
	\end{split}
\end{equation}
must be determined ($\hat{L}^{(T,D)}$ represents the version of $\hat{L}$ with dark-field, $\hat{L}^{(T)}$ the one without).
In the presented software, this minimization is performed in two steps.
Since the cost functions are only defined for integer values of $u_x$ and $u_y$,
the first step minimizes $\hat{L}^{(T,D)}$ or $\hat{L}^{(T)}$ only
on this grid of integer shifts, as described in subsection~\ref{sec:int_minimize}.
However, this step alone yields an insufficient precision for the phase signal ($u_x$
and $u_y$ rarely vary by more than one pixel for typical imaging setups),
necessitating a sub-pixel interpolation step of the cost function landscape,
described in subsection~\ref{sec:subpixel}.

\subsection{Discrete minimization step}\label{sec:int_minimize}
Starting at $\bu = \mathbf{0}$, the optimization procedure in \bu-space is done by successive one-dimensional downhill optimization subroutines, performed alternatingly in the $u_x$ and $u_y$ directions.
The one-dimensional optimization compares cost function values of the current estimate with its neighbors and varies the estimate by one pixel per step.
Previously calculated cost function values are cached in order to maximize computation speed.
An estimate is considered to be a minimum in one dimension if it is lower than both neighbor values. When this happens, an identical minimization procedure in the orthogonal direction begins, starting from the current estimate.
This procedure repeats until an estimate is simultaneously a minimum in both directions.
We call this value $\bu_\mathrm{d}$ the discrete minimum.

\begin{figure}[ht]
	\centering\includegraphics{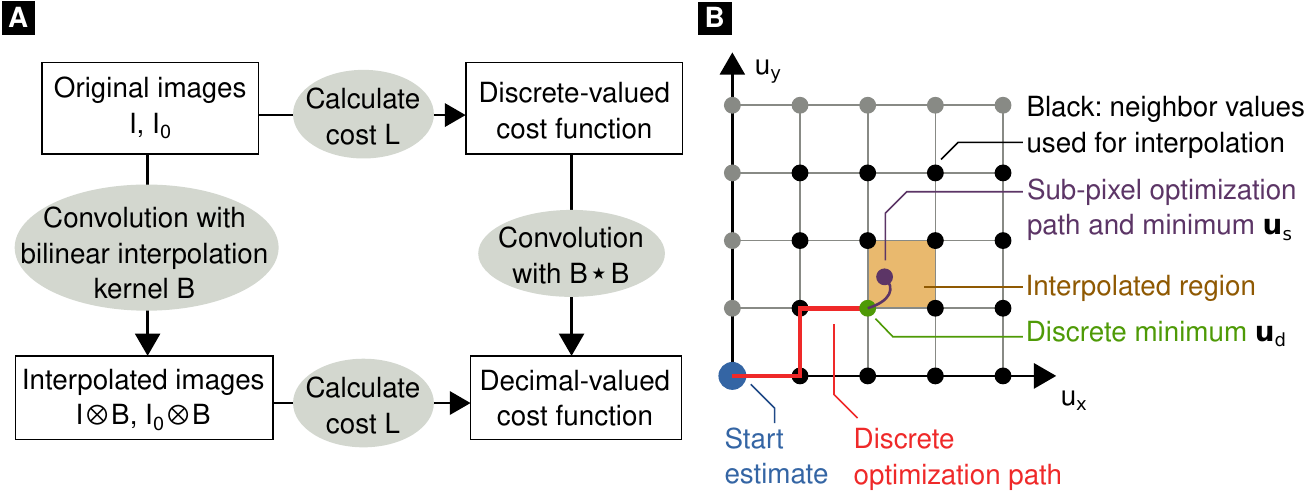}
	\caption{(a) Approach for cost function interpolation.
		Performing bilinear interpolation (convolution with the kernel $B$) on the
		input images, and then calculating the cost function of the interpolated
		values is (approximately) equivalent to directly calculating cost function
		values and then interpolating this with the kernel $B\star B$.
		(b) Procedure for identifying sub-pixel shift values. From the starting
		estimate ${\bu = (0,0)}$, optimization is first done on the grid of
		integer-valued shifts. Once the discrete minimum $\bu_\mathrm{d}$ is found, the convolution
		of the cost function with $B \star B$ is calculated in a $1\times 1$ pixel area adjacent to
		the minimum. The choice of ``quadrant'' relative to $\bu_\mathrm{d}$ is determined by the location of its neighbors with the
		lowest cost function values. Newton-Raphson optimization on the cost function in this area determines the sub-pixel optimum $\bu_\mathrm{s}$. Cost function values
		from a $4\times 4$ square of pixels surrounding the interpolated area (black dots)
		are required for the convolution due to the $4 \times 4$ pixel footprint of
		the interpolation kernel. Note that the estimates for $T$ and $D$
		are not modified after sub-pixel optimization.}
	\label{fig:bilinear_kernel}
\end{figure}

Since the refined, sub-pixel-precision minimum must lie within $\pm 1\,\text{pixel}$ of the discrete minimum, it is sufficient to interpolate a $1 \times 1$ pixel neighborhood for minimization.
It is computationally efficient to interpolate in a $1 \times 1$ square \emph{between} the points of the integer grid (see highlighted area in Fig.~\ref{fig:bilinear_kernel}b), since the interpolated surface in this area is fully defined by the smallest possible number of neighboring cost
function values, namely $4 \times 4 = 16$.

As four such areas (``quadrants'') are adjacent to the found minimum $\bu_\mathrm{d}$, it must be determined ahead of time which one contains the sub-pixel minimum.
This is done by comparing the cost function values of the vertical and horizontal neighbors of $\bu_\mathrm{d}$, and selecting the quadrant adjacent to the neighbors with the lowest values.

\subsection{Sub-pixel minimization step}\label{sec:subpixel}
A simple method to identify the cost function minimum with sub-pixel precision is to interpolate the recorded images $I_m$, $I_{0,m}$, and thus evaluate the resulting interpolation functions on a finer grid than the original images.
Calculating the cost function from these images then produces a more finely sampled version of the original cost function, thus yielding phase-shift values with greater accuracy.
We found that this process can be simplified by directly convolving the original cost function landscape with a specific kernel. This idea is schematically shown in Fig.~\ref{fig:bilinear_kernel}a.

Bilinear interpolation can be expressed as a convolution of the input function (defined on a grid) with an interpolation kernel $B$.
We show in Appendix~\ref{sec:appendix_int_kernel} that the cost function landscape calculated from bilinearly interpolated intensities is given by the convolution of the non-interpolated cost function landscape $\hat{L}(\br;\bu)$ with the kernel $B \star B$, i.e., the autocorrelation function of $B$.

Since this kernel has an analytical expression, the interpolated cost function surface and its local first and second positional derivatives do as well.
This motivates the use of the Newton-Raphson method to identify the location of the minimum: beginning at the integer minimum $\bu_\mathrm{d}$, each update of the current estimate requires calculation of the first and second partial derivatives of the interpolation of $\hat{L}$.
Optimization is terminated when the magnitude of the position update falls below a threshold ($10^{-4}\,\text{pixels}$ by default).

As the used interpolation kernel has a footprint of $4 \times 4$ pixels, correct calculation of interpolated costs in a $1 \times 1$ pixel square requires evaluation of the cost function in a $4 \times 4$ pixel neighborhood around $\bu_\mathrm{d}$ (see black dots and highlighted area in Fig.~\ref{fig:bilinear_kernel}b).
Reusing cached cost function values from the discrete minimization reduces the number of function evaluations necessary for this.
Note also that the estimates for $T$ and $D$ are determined during the discrete minimization step, and are not updated after sub-pixel minimization, as we consider this change negligible.

As illustrated in Fig.~\ref{fig:bilinear_kernel}b, identifying the sub-pixel minimum of the cost function $\hat{L}$ (i.e., $\hat{L}^{(T)}$ or $\hat{L}^{(T,D)}$) for one image pixel $\br$ is achieved by:
\begin{enumerate}[noitemsep]
	\item Identifying the integer shift $\bu_\mathrm{d}$ for which $\hat{L}$ is minimized and finding the ${1 \times 1}$~pixel ``quadrant'' with the lowest neighbors.
	\item Evaluating $\hat{L}$ for a $4 \times 4$ neighborhood of points (integer shifts) centered on this quadrant
	\item Convolution of this neighborhood with the interpolation kernel $B \star B$
	\item Newton-Raphson minimization of the interpolated cost function	landscape, yielding a sub-pixel minimum $\bu_\mathrm{s}$.
\end{enumerate}

\section{Improvements by the new UMPA implementation}\label{sec:improvements}

\subsection{Addition of the ``sample-stepping'' acquisition technique}\label{sec:diffuser_sample_stepping}

The UMPA model can combine image data with different relative positions of
diffuser and imaged sample (indexed by the subscript $m$ in the preceding equations).
There are important benefits for doing so:
\begin{itemize}[noitemsep]
	\item \emph{Improved convergence:} speckle patterns often look similar to many shifted versions of themselves.
	Thus, if only one diffuser position is used ($M=1$), the cost function may exhibit several local 	minima in $\bu$-space, which increases the probability of minimization converging to an incorrect solution.
	If multiple diffuser positions are combined, however, the full cost function landscape is given by 	the sum of landscapes for each position.
	Since the locations of incorrect local minima are mostly uncorrelated for sufficiently large 	diffuser displacements, they increasingly cancel out, while the correct minimum is reinforced.
	\item \emph{Spatial resolution:} while UMPA can be used with a single diffuser position ($M=1$), this is not compatible with the window size parameter $N=0$ (i.e., an analysis window of ${1 \times 1}$~pixel), since $3$ or $4$ unknowns would have to be retrieved from only $2$ data points.
	The use of analysis windows with $N>0$ is however always associated with a decrease in spatial resolution.
	Thus, pixel-size resolution is only achievable with the use of multiple diffuser steps, an approach similar to the original XSVT model~\cite{Berujon2015}.
	\item \emph{Noise performance:} naturally, combining multiple diffuser positions also increases total detector dose and thus the contrast-to-noise ratio in all retrieved image modalities.
\end{itemize}

\begin{figure}[ht]
	\centering\includegraphics{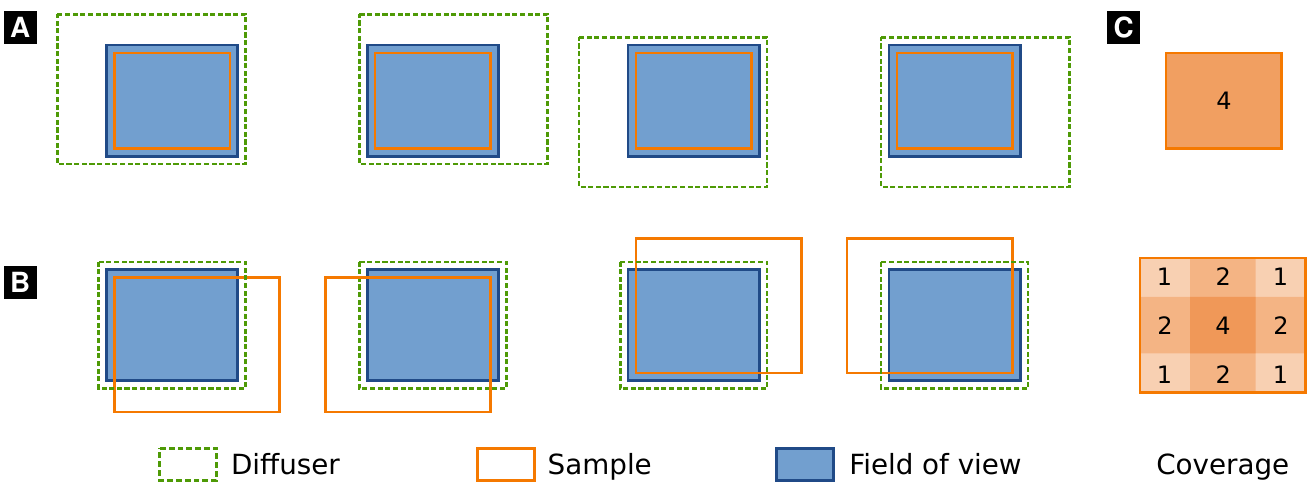}
	\caption{Illustration of the two available stepping methods. (a)~\emph{Diffuser stepping}: the 
		diffuser is laterally translated, while the sample remains stationary.
		(b)~\emph{Sample stepping}: the diffuser remains stationary and the sample is translated.
		(c)~``Coverage maps'' of the sample for the examples in (a) and (b), showing how many times 
		each part of the sample is imaged.
		Sample stepping increases the size of the covered region, at the expense of a locally decreased 
		coverage.}
	\label{fig:stepping_methods}
\end{figure}

\begin{figure}[ht]
	\centering
	\includegraphics{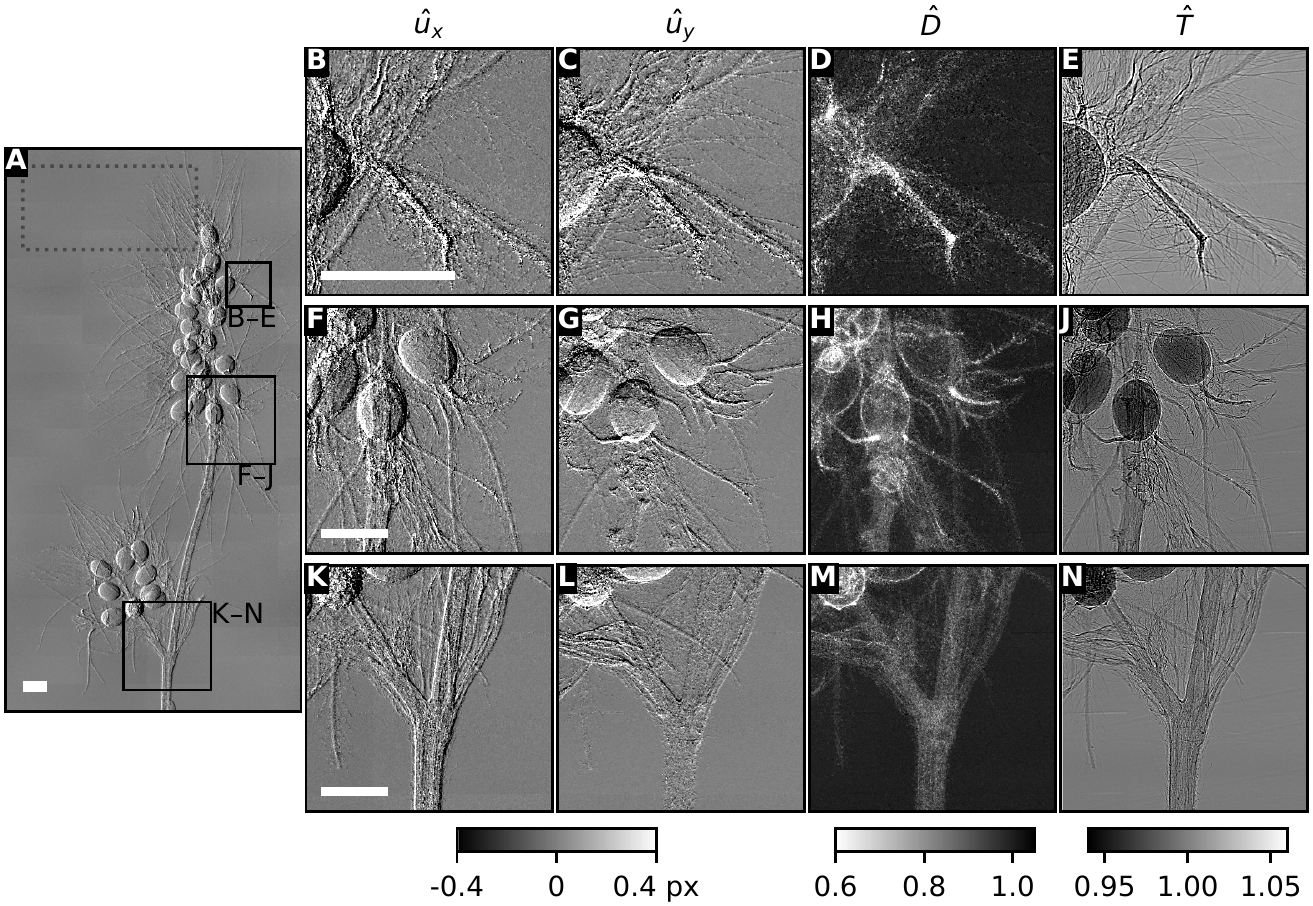}
	\caption{
		Flower (Hare's foot clover) imaged at the P05 beamline, PETRA~III, DESY, Hamburg with the 
		sample-stepping technique.
		(a):~Overview of the whole sample ($\hat{u}_x$). The dashed rectangle represents the size of the detector 
		field of view ($3.48\,\mathrm{mm} \times 7.25\,\mathrm{mm}$). Three regions of interest 
		(solid rectangles) are shown in the insets on the right, for all four modalities (columns from left
		to right: horizontal shift $\hat{u}_x$, vertical shift $\hat{u}_y$, dark-field $\hat{D}$, 
		transmittance $\hat{T}$). The images are processed with the window size parameter $N=2$.
		The sample was translated on a rectangular grid of $25 \times 10$ positions. $E=35\,\mathrm{keV}$, pixel size: \SI{0.916}{\micro\meter}. Sample-detector distance: $20\,\mathrm{cm}$, diffuser ($9 \times 400$-grit sandpaper) $11.5\,\mathrm{cm}$ 
		upstream of sample. Scale bar (white): $1\,\mathrm{mm}$.
		The original size of (a) is $12.4\,\mathrm{mm} \times 23.5\,\mathrm{mm}$, i.e., $13500 \times 
		25700$~pixels, but the shown data is binned [$10 \times 10$ in~(a), $3 \times 3$ in~(b)--(n)].}
	\label{fig:flower}
\end{figure}

\begin{figure}[ht]
	\centering
	\includegraphics{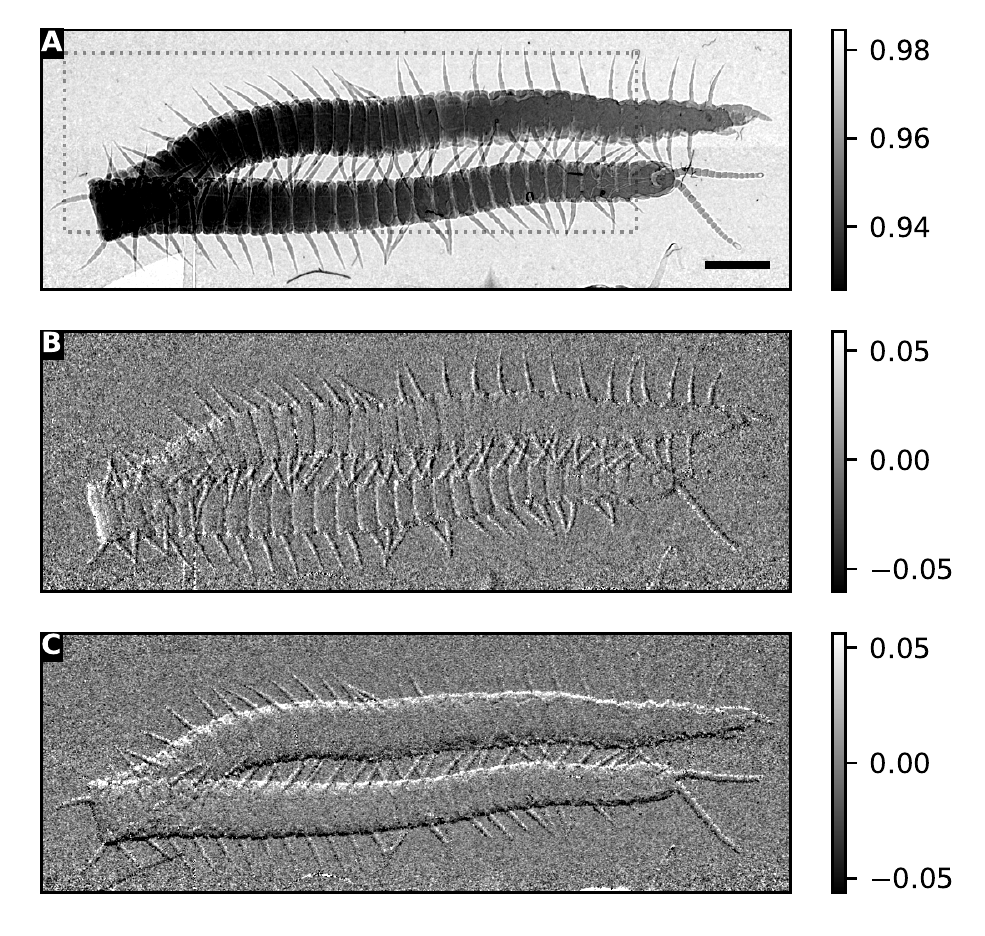}
	\caption{
		Centipede imaged at the SYRMEP beamline (Elettra Sincrotrone, Trieste, Italy) with the 
		sample-stepping technique.
		(a): transmittance $\hat{T}$, (b, c): horizontal and vertical refraction $\hat{u}_x$, 
		$\hat{u}_y$ (in pixels). The dark-field modality is omitted here due to a lack of visible features.
		The dashed rectangle represents the detector field of view ($5.46\,\mathrm{mm} \times 
		17.46\,\mathrm{mm}$).
		The sample was translated on a rectangular grid of $M=2 \times 26$ positions, $N=2$.
		$E=20\,\mathrm{keV}$, sample-detector distance $32\,\mathrm{cm}$, diffuser ($5\times320$-grit sandpaper) $111.2\,\mathrm{cm}$ upstream of sample.
		Eff. pixel size: \SI{8.8}{\micro\meter}. Scale bar is $2\,\mathrm{mm}$.}
	\label{fig:centipede}
\end{figure}

The previously established method to achieve these diffuser-to-sample displacements is to translate 
the diffuser while the sample is kept stationary.
Here we call this approach \emph{diffuser stepping}.
However, since only the relative shift between sample and diffuser is crucial, it is also possible to 
laterally displace the sample, and then modify the mathematical model to take this displacement into 
account.
We call this \emph{sample stepping}.
The difference between the two methods is illustrated in Fig.~\ref{fig:stepping_methods}, and 
example datasets acquired with sample stepping are shown in Fig.~\ref{fig:flower} and 
Fig.~\ref{fig:centipede}.
The cost functions are modified to account for sample translations by subtracting the sample 
displacement $\sm$ in each frame, as shown in section~\ref{sec:math}.

Here, $\mathbf{S}_m$ is a 2D vector describing the relative transverse position of the sample in the 
$m$-th frame, in multiples of the effective pixel size, and relative to the starting point (e.g. the first 
frame).
For diffuser stepping, $\sm$ is simply $\mathbf{0}$ for all $m$.
Conversely, the location of the diffuser is not explicitly required for signal retrieval (neither for 
sample stepping nor for diffuser stepping), and thus not expressed in the mathematical model 
presented here.

Fig.~\ref{fig:stepping_methods}c schematically shows ``coverage maps'' achieved with the
different stepping approaches, i.e. the map of the number of times each part of the
sample was part of an acquisition. Diffuser stepping always yields constant
coverage across the field of view (FOV), while the coverage of sample-stepping measurements
encompasses a larger area, but gradually decreases towards the edge of
this area.
This leads to a decrease of image quality in these regions.
Besides this basic trade-off, these are some of the main benefits of using the
sample stepping approach:
\begin{itemize}[noitemsep]
	\item \emph{Faster acquisition of very large samples:} acquiring an object far
	larger than the field of view with diffuser stepping generally requires a larger
	number of motor movements (doing a full diffuser stepping, moving on to the next 
	sample position, etc.), and thus requires more overhead time.
	\item \emph{High-throughput imaging:} for low- or medium-resolution applications,
	sample stepping could be performed with continuous sample displacement, e.g., via
	a conveyor belt. This would eliminate motor displacement overhead and could lead
	to drastic speed-ups in acquisition time, similar to ``fringe-scanning'' used in
	grating-based X-ray~\cite{Kottler2007}.
	\item \emph{Simpler instrumentation:} only one motorized stage is required (the
	one for the sample). Furthermore, no high repeatability is required for this
	stage, since each position is approached only once. An accurate readout of this
	stage's position is still useful, but deviations can be corrected through
	cross-correlation analysis.
	
\end{itemize}
However, the approach may also introduce some (situational) drawbacks:
\begin{itemize}[noitemsep]
	\item \emph{Drifts:} for longer acquisitions with many frames (e.g., a large
	two-dimensional grid of sample positions), drifts of sample and/or diffuser
	positions may occur, leading to artifacts. However, this can be ameliorated
	by frequently acquiring reference images, and deviations can often be
	corrected with cross-correlation analyses.
	\item \emph{Identifying sample limits:} determining the extent of a
	large object in terms of sample motor positions can be difficult, especially
	when the reduction of coverage towards the edge of the reconstructed area
	is taken into account (cf. Fig.~\ref{fig:stepping_methods}c).
	The sample may then hardly be visible for the ``edge'' sample positions.
	For samples smaller than the detector FOV, care should be taken
	that the object does not leave the FOV for any of the motor positions.
	\item \emph{Effective pixel size must be known / determined:} this is
	necessary since the displacement vectors $\sm$ must be expressed as multiples
	of the pixel size. However, performing UMPA reconstruction with different
	pixel size estimates and comparing the quality of output images yields quick and
	precise results.
	\item \emph{Variable image quality / noise levels due to the non-constant
		coverage.} For achieving roughly constant coverage in the center of the
	reconstructed regions, we recommend the use of a one- or two-dimensional grid
	of motor displacements, equal to an integer fraction of the field of view,
	i.e.: $\Delta x = W_x / N_x$, $\Delta y = W_y / N_y$, where $W_x$ and $W_y$
	is the extent of the detector FOV in $x$ and $y$, and $N_x, N_y \in \mathbb{N}$.
	This yields a maximum coverage of $N_x$ or $N_y$ for 1D stepping, and $N_x N_y$
	for a 2D stepping grid. Furthermore, a helper function is available in the
	software package to plot coverage maps for a given sample trajectory.
	\item \emph{Limited compatibility with cone-beam geometry:} in setups with a high ratio of detector FOV and source-to-detector distance, sample stepping may produce artifacts due to projection inconsistencies between different sample displacements, especially for thick samples.
\end{itemize}

\subsection{Speedup and software architecture}\label{sec:benchmarks}

C++ is a very widely used programming language that combines high computational speed with the ability to use object-oriented programming concepts, thus making it suitable for complex 
applications where high efficiency is required.
Due to its extensive use, C++ compilers exist that can generate highly optimized code for nearly all computing architectures and operating systems.
However, these advantages come at a cost: C++ code is often more complex, less readable, and thus more difficult to modify than an equivalent implementation in an interpreted language such as Python.
Due to its high flexibility and ease of use, Python has been widely adopted by the scientific computing community, and a large ecosystem of Python modules for numerical and scientific applications has since emerged. 

In order to bridge the gap between the worlds of Python and C/C++, the Cython language has been developed~\cite{Behnel2011}. It features a syntax similar to Python, but is compiled into C or C++ code (and then into machine code by a C/C++ compiler).
This enables the two main use cases of Cython: writing high-performance code with a readability and flexibility similar to Python, and integrating existing C/C++ modules in a Python interface.
Both aspects were used for the work presented here in order to combine the high efficiency of C++ with the accessibility of Python.

Due to the language's implementation, Python programs are usually limited to a single process. 
This is of crucial importance for UMPA, because its central problem, the minimization of the cost function, is ``embarrassingly parallel'', i.e., it is easy to parallelize with minimal overhead, since the problem is independent for each image pixel.
Parallelization in Python can still be achieved by spawning multiple processes, but this is slower than multithreading, has greater overhead, and is more prone to complications.
With Cython however, multiprocessing can be achieved easily and with minimal overhead via OpenMP.
The speedup achieved by the use of C++ and multithreaded processing are illustrated in Table~\ref{tab:benchmarks}.

The core element of the presented software package is the implementation of the cost function calculation, and its minimization, in C++.
A Cython module provides the interface between these C++ libraries and the Python front-end.
A class-oriented program structure is used both in the C++ and Cython sections of the module, which avoids repetition of code between the two available UMPA models (with and without dark-field).
A code example with an explanation of the individual steps is provided in Appendix~\ref{sec:appendix_code_example}.
\begin{table}[!ht]
	\caption{Runtime measurements of the Python and C++ implementations of UMPA. Benchmarks 
	were run on a compute server with $512\,\mathrm{GByte}$ of memory and two AMD~EPYC~7551 
	CPUs with $32$ cores each. Runtime values for the C++/Cython benchmarks are given as mean 
	and standard deviation of five runs.}
	\label{tab:benchmarks}
	\begin{tabular}{lllllll}
		Image size [px] & $M$ & $N$ & UMPA version & Threads &  Runtime & Rel. speed-up \\
		\hline
		\multirow{4}{*}{$1000 \times 1000$} & \multirow{4}{*}{$25$} & 
		\multirow{4}{*}{$2$} & Python & $1$ & $3864\,\mathrm{s}$ & $1$ \\
		& & & C++/Cython & $1$  & $(30.90 \pm 0.03)\,\mathrm{s}$ & $125$\\
		& & & C++/Cython & $16$ & $(2.10 \pm 0.02)\,\mathrm{s}$  & $125\times 14.7$\\
		& & & C++/Cython & $64$ & $(0.89 \pm 0.05)\,\mathrm{s}$  & $125 \times 34.7$ \\
		\hline
	\end{tabular}
	\vspace*{-4pt}
\end{table}

As illustrated in Table~\ref{tab:benchmarks}, the new C++ implementation of UMPA running on a 
single thread is about $125$ times faster than the Python implementation, while multithreading adds 
an additional speed-up factor closely related to the number of used threads.

\subsection{Correction of differential-phase estimation bias}\label{sec:bias}
The presented sub-pixel interpolation approach introduces a bias to the determined shifts $u_x$, 
$u_y$.
This is most obvious in sample-free areas of the FOV, or when comparing a set of images with itself, i.e., running UMPA with $I_{0,m} = I_m$.
Although the resulting cost function values for integer shifts have a minimum at $\bu = 
\mathbf{0}$, the determined sub-pixel minimum often has an offset.

On the other hand, explicitly oversampling a set of images ($I_{0,m}$, $m=1,2,\ldots$) and 
calculating the cost function of the image set with itself must still yield a minimum precisely at 
$\bu = \mathbf{0}$: a sufficiently ``random'' pattern should be most similar to an unshifted copy of itself.

This apparent contradiction is probably resolved by the fact that the interpolation kernel is not quite equivalent to oversampling of the raw data:
The interpolation step is exact for the cost function terms $l_1, \ldots, l_6$, and thus also for any 
linear combination of these terms, such as $L^{(T)}$ and $L^{(T,D)}$
[see Eq.~\eqref{eq:LT_short}, \eqref{eq:cost_terms_nodf}, \eqref{eq:LTD} in Appendix~\ref{sec:appendix_partial_opt}].
However, the minimization of these cost functions with respect to $T$, or $T$ and $D$ respectively, introduces non-linear operations [division and multiplication of the $l_j$ terms, see 
Eq.~\eqref{eq:LT_short_final}, \eqref{eq:alpha_beta_matrix}, \eqref{eq:TD_alpha_beta} in Appendix~\ref{sec:appendix_partial_opt}].
These operations do not commute with oversampling, which probably causes the observed bias.

\begin{figure}[ht]
	\centering
	\includegraphics{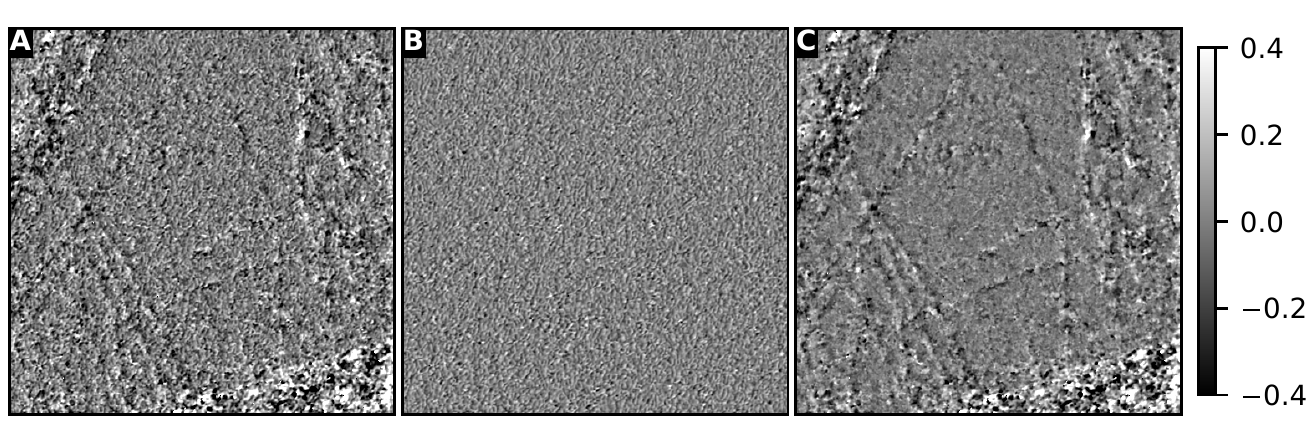}
	\caption{Illustration of bias correction. (a):~horizontal shift $\hat{u}_x$ from UMPA, 
	(b):~bias estimate $\hat{u}_{x}^\text{(ref)}$ obtained by matching reference images with 
	themselves, (c):~difference between (a) and (b). The fixed-pattern noise is clearly reduced, 
	yielding an improved contrast for faint structures (here: zoomed region of the flower shown in 
	Fig.~\ref{fig:flower}).}
	\label{fig:bias_example}
\end{figure}

Although the exact origin of the effect has not been fully ascertained, we found that the lateral shifts of the reference data set $I_0$ \emph{with itself} is a good estimate for this bias.
We thus use the following procedure to reduce the estimation bias of $\hat{\bu}$:
\begin{enumerate}[noitemsep]
	\item Calculation of $\hat{\bu}(\br)$ (including sub-pixel interpolation) with $I_m$ as sample data and $I_{0,m}$ as reference data
	\item Repeating the same calculation, but with $I_{0,m}$ as \emph{both sample data and 
	reference data}, i.e. substitution of $I_m$ with $I_{0,m}$, yielding the shifts 
	$\hat{\bu}^\text{(ref)}(\br)$
	\item Calculation of $\hat{\bu}^\text{(corr)}(\br) = \hat{\bu}(\br)$ -- $\hat{\bu}^\text{(ref)}(\br)$
\end{enumerate}
An example for this correction is shown in Fig.~\ref{fig:bias_example}. This corrects the majority of the bias for weak refraction (i.e., significantly less than $1\,\text{px}$).
For larger amounts, the effectiveness of this correction decreases.
We have found that the bias correction remains more accurate if the correction
is taken at a different pixel, shifted by $\hat{\bu}$:
\begin{equation}\label{eq:biascorr2}
	\begin{split}
		\hat{\bu}^\text{(corr)}(\br) = \hat{\bu}(\br) - \hat{\bu}^\text{(ref)}(\br+\hat{\bu}).
	\end{split}
\end{equation}

\subsection{Comparison of coordinate assignment schemes}\label{sec:shift_mode}

\begin{figure}[ht]
	\centering\includegraphics{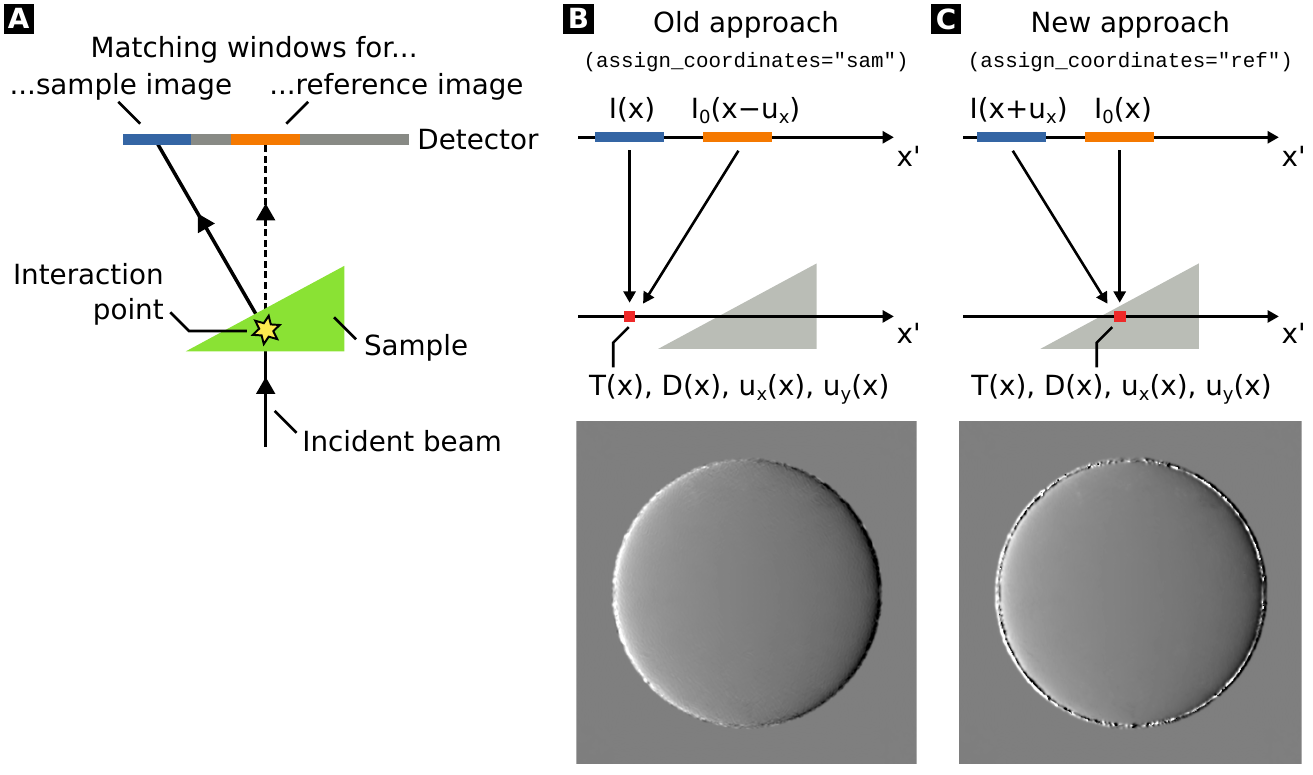}
	\caption{Illustration of the two coordinate assignment schemes. 
		(a)~Refraction by the sample induces a shift between matching analysis window pairs in the 
		sample and reference images, prompting the question of which position the algorithm's output 
		should be assigned to. (b)~For the old approach (\texttt{assign\_coordinates="sam"}), the calculated image data (i.e., transmittance, differential phase and dark-field) is assigned to the sample window center.
		(c)~For the new approach (\texttt{assign\_coordinates="ref"}), the values are instead assigned to the reference window center.
		Intuitively, this is more reasonable since the assignment point coincides with the interaction point of sample and incident light, but the shown example image shows that it introduces artifacts near sample edges.
		We found that this is due to UMPA avoiding sample regions with strong distortions of the speckle pattern.}
	\label{fig:shift_mode}
\end{figure}

The cost function introduced in section~\ref{sec:math} depends on an analysis window on the 
sample image(s), centered on some coordinate $\br'=\br - \mathbf{S}_m$, and a second window on 
the reference image(s), shifted by an amount $\bu$, i.e. centered on $\br' -\bu$.
This notation implies that the reference window moves during the minimization procedure (since 
$\bu$ is the optimized variable), and the sample window remains fixed.
In particular, the values returned by the minimization procedure ($\hat{T}$, $\hat{D}$, and 
$\hat{\bu}$) must then be assigned to $\br'$, the center of the sample window.
This is illustrated in Fig.~\ref{fig:shift_mode}b.
If the values were instead assigned to the center of the \emph{moving} window, there is a possibility 
that some pixels of the output images are never being assigned a value, and other pixels being 
assigned a value multiple times, since the UMPA optimization loop is run exactly once for every 
value of $\br'$.

However, Fig.~\ref{fig:shift_mode}a illustrates that it is more physically accurate to assign the UMPA output to the center of the reference window, since this point coincides more closely with the point of interaction between radiation and object.
This also corresponds with a modification suggested in~\cite{Morgan2020b} [Eq.~(20)].
Therefore, we have introduced an alternative mode for the optimization procedure where the reference window remains fixed and the sample window is moved during the optimization.
This is simply achieved by adding $\bu$ to all coordinates, i.e. the reference window is now centered on $\br'$, and the sample window is centered on $\br' + \bu$ (see Fig.~\ref{fig:shift_mode}c).
In the code, either minimization procedure can be selected with the parameter \texttt{assign\_coordinates} (see line~10 in the code example, Appendix~\ref{sec:appendix_code_example}).
In this case, the UMPA model [Eq.~\eqref{eq:UMPA_model} and \eqref{eq:cost_function_simple}] 
changes to:
\begin{equation}
	\begin{split}
		I^\text{(model)}(\br; T, D) &= T \left\{ D \left[ I_{0}(\br) - \langle I_{0} \rangle(\br) \right]
		+ \langle I_{0} \rangle(\br) \right\}, \\
		L(\br; \bu, T, D) &= \sum_{m=1}^{M} \sum_{w_x, w_y=-N}^{N} \!\!\!\!\!\! \Gamma(\bw) \left[ 
		I_m^\text{(model)}(\br+\bw-\mathbf{S}_m, T, D) - I_m(\br+\bu+\bw-\mathbf{S}_m) \right]^2.
		\label{eq:cost_function_D_shift_true}
	\end{split}
\end{equation}

However, the example images in Fig.~\ref{fig:shift_mode} (generated from simulated speckle data of a sphere) show that this new approach performs poorly near sample edges.
This is due to a distortion of speckle patterns in the presence of sample edges (e.g., due to strong wavefront curvature or propagation fringes), thus exhibiting low similarity to all regions in the reference speckle pattern.

For the new approach (\texttt{assign\_coordinates="ref"}, see lines~10 and 14 in the code example), this leads to a cost function increase as the sample window approaches an edge, and the algorithm thus avoids selecting such sample window positions, leading to the black-and-white ring artifact in Fig.~\ref{fig:shift_mode}c.
For the old approach (\texttt{assign\_coordinates="sam"}) however, the optimization is performed with a stationary sample window.
In the case of a sample window centered on an edge, the cost function increase due to the distortion of the pattern is present for all shifts $\bu$ (i.e., all possible reference window matches), and thus essentially behaves as an offset to the cost function landscape, having a much weaker impact on the minimization procedure.

The old approach is therefore maintained as the default method in the present implementation.
Since the two methods essentially minimize different subsets of the same cost function space, we believe it is possible to synthesize both approaches into a method which is both physically accurate, and numerically robust.
One option might be to process with the old approach, and then correct the misattribution of positions by applying an image warping transform to the images, using a warp map generated from the found shift values $\hat{\bu}(\br)$ [or the bias-corrected $\hat{\bu}^\text{(corr)}(\br)$].
This being said, we believe that the magnitude of this misattribution is minimal for most imaging tasks, since the vast majority of retrieved differential phase shifts are below one pixel.


\subsection{Addition of options for weighting and region-of-interest processing}\label{sec:masks}
The subset of data to include in the UMPA reconstruction can be reduced arbitrarily through the use of a pixel mask (see parameter \texttt{mask\_list} in the code example, Appendix~\ref{sec:appendix_code_example}).
This mask has the same dimensions as the input data, so that the contribution of each raw data pixel can be individually controlled.
This is especially useful if only some of the raw image frames in a dataset contain artifacts.
A typical use case for masks is with an iterative workflow, i.e.: processing the full data set, identifying image regions containing artifacts, excluding the matching data subsets in the mask, and re-processing.
Alternatively, mask values can be set to non-boolean values and will then be interpreted as a weighting for the UMPA cost function.

If sample stepping is used and a region to be excluded is stationary relative to the sample, formulating a suitable mask to exclude it in all frames may be difficult, since this region must be tracked with the sample's movement (cf. Fig.~\ref{fig:stepping_methods}b).
In this case, the \texttt{ROI} feature can be useful: it functions similarly to the ``mask'' feature, but coordinates are given in the reference frame of the reconstructed image, not the raw data.
The \texttt{ROI} parameter is given as a pair of NumPy slice objects or \texttt{(start, stop, step)} pixel index ranges.
It can thus be used to reduce the reconstructed field of view, or to reduce the resolution of the reconstruction by calculating only every \texttt{step}-th pixel value.
The latter is especially useful if very large analysis windows are used: since the resolution of output data is restricted by the window size, the \texttt{step} parameter can be safely increased without loss of information.

\section{Conclusion and Outlook}\label{sec:conclusion}
We presented an improved implementation of the "Unified Modulated Pattern Analysis" algorithm.
Crucial implementation details, such as the definition of the cost function and its minimization, were discussed in-depth.
The software is available in a public GitHub repository (see the ``Data Availability'' section below).
It includes support for the ``sample stepping'' technique, which obviates the need for motorized displacement of the structure generating the modulated intensity pattern, and allows imaging samples far larger than the detector field-of-view.
We demonstrated the feasibility of this technique with several examples.

As the software is far more computationally efficient than the preceding Python implementation, and is also capable of multithreading, it is ideally suited for processing large volumes of wavefront-marking X-ray imaging data, such as tomography datasets or large-field-of-view sample-stepping projection images.
Even the combination of both techniques, ``sample-stepping tomography'', where the tomography axis is laterally displaced between tomographic scans, is feasible.
The presented software has already been successfully used in published work with two-dimensional gratings~\cite{Gustschin2021} and sandpaper~\cite{Smith2022} as wavefront-marking devices.

We discussed the presence of an estimation bias in the produced differential-phase images and showed a simple approach to reduce this effect.
Finally, we discussed the inherent ambiguity in the assignment of fit results to pixel coordinates.
We compared a newly introduced method to resolve this problem with the older, previously used approach and found that, although the new approach is in principle more physically accurate, it produces image artifacts of greater magnitude near sample edges.

Besides this, the physical model underlying UMPA can be further refined:
in particular, the mathematical model used for the dark-field signal is an imperfect description of the underlying physics, since all spatial frequencies of the speckle pattern (except zero) are scaled by the same factor.
Instead, it is more realistic to characterize small-angle scatter by convolution with a spatially variable Gaussian blur kernel.
The adaptation of UMPA to such a model, and its successful application to measurements of highly directional carbon fiber samples has recently been demonstrated~\cite{Smith2022}.

\appendix
\counterwithin*{equation}{section}
\counterwithin*{figure}{section}
\renewcommand{\theequation}{\thesection\arabic{equation}}
\renewcommand{\thefigure}{\thesection\arabic{figure}}
\section{Partial optimization of the UMPA cost function}
\label{sec:appendix_partial_opt}
This section contains the mathematical details for minimizing the two versions of the UMPA cost function (with and without dark-field) with respect to $T$, and both $T$ and $D$, respectively. This is an intermediate step in the procedure of global cost function minimization.

\subsection{Partial optimization for the model without dark-field}\label{sec:partial_opt_T}

When setting $D=1$ in Eq.~\eqref{eq:UMPA_model}, 
the cost function in Eq.~\eqref{eq:cost_function_simple} 
can be rewritten as
\begin{equation}
	\begin{split}\label{eq:LT_short}
		L^{(T)} = T^2 \cdot l_3 -2T \cdot l_5 + l_1,
	\end{split}
\end{equation}
where
\begin{equation}\label{eq:cost_terms_nodf}
	\begin{split}
		l_1 &= \sum_{m,\bw} \Gamma(\bw) I_m^2(\br+\bw-\mathbf{S}_m),\\
		l_3 &= \sum_{m,\bw} \Gamma(\bw) I_{0,m}^2(\br+\bw-\mathbf{S}_m-\bu),\\
		l_5 &= \sum_{m,\bw} \Gamma(\bw) I_{0,m}(\br+\bw-\mathbf{S}_m-\bu) 
		I_m(\br+\bw-\mathbf{S}_m).\\
	\end{split}
\end{equation}
Since, for a local minimum, $\partial L_T / \partial T = 2T \cdot l_3 - 2 l_5=0$,
\begin{equation}\label{eq:T_short}
	\begin{split}
		\hat{T} = l_5 / l_3.
	\end{split}
\end{equation}
Reinserting Eq.~\eqref{eq:T_short} into Eq.~\eqref{eq:LT_short}, this yields
\begin{equation}
	\begin{split}\label{eq:LT_short_final}
		\hat{L}^{(T)} = l_1 - l_5^2 / l_3.
	\end{split}
\end{equation}
Thus, for each estimate of $\bu$, the terms $l_1$, $l_3$, and $l_5$ are calculated, and from these, the cost and the transmittance estimates are calculated according to Eq.~\eqref{eq:T_short}, \eqref{eq:LT_short_final}.
An equivalent calculation can be performed for the model including dark-field, as shown below.

\subsection{Partial optimization for the model with dark-field}\label{sec:partial_opt_TD}

Here we show the equivalent of the calculation in subsection~\ref{sec:partial_opt_T} for the model 
including dark-field.
We substitute the fit variables $T$, $D$ by the quantities $\alpha=TD$, $\beta=T(1-D)$ and different summation terms by $l_1, \ldots, l_6$.
\begin{equation}
	\begin{split}\label{eq:LTD}
		L^{(T,D)}(\br; \bu, T, D) &= l_1 +\beta^{2} l_2 + \alpha^{2} l_{3} - 2\beta l_{4} -2\alpha 
		l_{5} + 2\alpha\beta l_{6},\\
		\text{where} \qquad l_1 &= \sum_{m, \bw} \Gamma(\bw) I_m^2(\br + \bw - \sm), \\
		l_2 &= \sum_{m, \bw} \Gamma(\bw) {\langle I_{0,m} \rangle}^2 (\br + \bw - \sm - \bu), \\
		l_3 &= \sum_{m, \bw} \Gamma(\bw) I_{0,m}^2 (\br + \bw - \sm - \bu), \\
		l_4 &= \sum_{m, \bw} \Gamma(\bw) \langle I_{0,m} \rangle (\br + \bw - \sm - \bu) I_m(\br + 
		\bw - \sm), \\
		l_5 &= \sum_{m, \bw} \Gamma(\bw) I_{0,m} (\br + \bw - \sm - \bu) I_m(\br + \bw - \sm), \\
		l_6 &= \sum_{m, \bw} \Gamma(\bw) \langle I_{0,m} \rangle (\br + \bw - \sm - \bu) I_{0,m} (\br 
		+ \bw - \sm - \bu).
	\end{split}
\end{equation}

The terms $l_i$ depend on $\br$, as well as on $\bu$. Note that $l_1$, $l_3$ and $l_5$ are identical 
to the ones used by the model without dark-field.
These six terms are thus calculated for each new estimate of shift $\bu$.
As before, we know that for a local minimum of $L^{(T,D)}$, its first derivative with respect to $T$ 
and $D$, and thus also with respect to $\alpha$ and $\beta$, is zero:
\begin{equation}
	\begin{split}
		\left.\frac{\partial L^{(T,D)}}{\partial \alpha}\right|_{\substack{\alpha=\hat{\alpha},\\ 
				\beta=\hat{\beta}}}
		= 2 \hat{\alpha} l_3 - 2 l_5 + 2 \hat{\beta} l_6 = 0,
		\left.\frac{\partial L^{(T,D)}}{\partial \beta}\right|_{\substack{\alpha=\hat{\alpha},\\ 
				\beta=\hat{\beta}}}
		= 2 \hat{\beta} l_2 - 2 l_4 + 2 \hat{\alpha} l_6 = 0.
	\end{split}
\end{equation}

\noindent This can be expressed as a matrix multiplication, which also reveals the solution:
\begin{equation}\label{eq:alpha_beta_matrix}
	\begin{split}
		\begin{pmatrix}
			l_3 & l_6\\ l_6 & l_2
		\end{pmatrix}
		\begin{pmatrix}
			\hat{\alpha} \\ \hat{\beta}
		\end{pmatrix}
		=
		\begin{pmatrix}
			l_5 \\ l_4
		\end{pmatrix}
		\Rightarrow
		\begin{pmatrix}
			\hat{\alpha} \\ \hat{\beta}
		\end{pmatrix}
		=
		{\begin{pmatrix}
				l_3 & l_6\\ l_6 & l_2
		\end{pmatrix}}^{-1}
		\begin{pmatrix}
			l_5 \\ l_4
		\end{pmatrix}
		=
		\frac{1}{l_3 l_2 - l_6^2}
		\begin{pmatrix}
			l_2 l_5 - l_4 l_6 \\ l_3 l_4 - l_5 l_6
		\end{pmatrix}.
	\end{split}
\end{equation}
Finally, we can retrieve $T$ and $D$ from $\alpha$ and $\beta$ via:
\begin{equation}\label{eq:TD_alpha_beta}
	\begin{split}
		\hat{T} = \hat{\alpha} + \hat{\beta},\ \hat{D}= \hat{\alpha} / (\hat{\alpha} + \hat{\beta}).
	\end{split}
\end{equation}
Inserting these values for $T$ and $D$ in Eq.~\eqref{eq:LTD} yields the cost function value
as a function of only the differential shifts:
\begin{equation}\label{eq:ltd_final}
	\begin{split}
		\hat{L}^{(T,D)}(\br;\bu) = L^{(T,D)}(\br; \bu, \hat{T}, \hat{D}).
	\end{split}
\end{equation}

\section{UMPA code example}
\label{sec:appendix_code_example}
Below is a use example for the new UMPA implementation, followed by an explanation of individual 
steps:

\begin{lstlisting}[language=Python]
	import numpy as np
	import UMPA as u
	sample = np.load("/path/to/sample.npy")
	ref    = np.load("/path/to/ref.npy")
	grid   = np.load("/path/to/grid.npy")
	mask   = np.load("/path/to/mask.npy")
	roi    = np.s_[50:-50:2,50:-50:2]
	m = u.model.UMPAModelDF(sample, ref, pos_list=grid, mask_list=mask,
	window_size=4, max_shift=5, ROI=roi)
	m.assign_coordinates = "sam"
	res = m.match(num_threads=16)
	mb = u.model.UMPAModelDF(ref, ref, pos_list=grid, mask_list=mask,
	window_size=4, max_shift=5, ROI=roi)
	mb.assign_coordinates = "sam"
	resb = mb.match(num_threads=16)
	ux, uy = res["dx"] - resb["dx"], res["dy"] - resb["dy"]
\end{lstlisting}

The variable \texttt{sample} holds a stack of images (NumPy arrays) with diffuser and sample, 
while \texttt{ref} holds the image stack with only the diffuser in the beam.
These image volumes should be contiguous in memory and in the double-precision floating-point 
(\texttt{numpy.float64}) data type so that they are correctly interpreted by the C++ routines.
However, the module can also easily be compiled to use the single-precision floating point 
(\texttt{numpy.float32}) data type instead. 
The \texttt{grid} variable (used below for the parameter \texttt{pos\_list}) is only required for 
sample-stepping measurements, where the sample is moved laterally instead of the diffuser 
(introduced in subsection~\ref{sec:diffuser_sample_stepping}), 
and contains the $(x, y)$ positions of the sample motor stage, in multiples of the effective pixel size.
The \texttt{mask} variable (used as parameter \texttt{mask\_list}) can optionally be used to 
selectively exclude data (e.g., bad pixels) for use with UMPA. It must be a NumPy array of the same 
shape as \texttt{sample} and \texttt{ref}. If it holds non-binary values, it is instead used as an 
additional per-pixel weighting factor for the UMPA cost function.
The \texttt{ROI} parameter is also optional and can be used to restrict the range of processed pixel 
values. While the \texttt{mask\_list} parameter is applied relative to detector coordinates, 
\texttt{ROI} is applied relative to sample coordinates, which is an important distinction when the 
sample-stepping mode is used.
These two parameters are discussed in greater detail in subsection~\ref{sec:masks}. 

Lines~8--9 create a model object which holds references to the image data, as well as processing 
parameters.
Two models are available: \texttt{UMPAModelDF} includes dark-field, and 
\texttt{UMPAModelNoDF} does not.
The parameter \texttt{window\_size} represents the variable $N$ from sections~\ref{sec:math} and \ref{sec:minimization}, 
while \texttt{max\_shift} is an upper threshold for the absolute value of the 
shifts $(u_x, u_y)$ before the discrete minimization is interrupted.

Line~10 sets the value of the \texttt{assign\_coordinates} parameter, which determines the position 
to which a determined match between sample and reference analysis window is assigned:
For \texttt{assign\_coordinates="ref"}, it is assigned to the center of the reference analysis window, 
while for \texttt{assign\_coordinates="sam"}, it is assigned to the center of the sample analysis 
window.
This is discussed in more detail in subsection~\ref{sec:shift_mode}. 

Finally, the actual minimization procedure is executed in line~11, with the number of threads being 
controlled by the \texttt{num\_threads} parameter. Upon completion, the \texttt{res} variable holds 
a Python dictionary containing the four image modalities (attenuation $T$, horizontal and vertical 
analysis window shifts $(u_x, u_y)$ in pixels, and dark-field $D$), as well as the minimized cost 
function value [$\hat{L}^{(T,D)}(\br; \bu)$ or $\hat{L}^{(T)}(\br; \bu)$] for each pixel.

Lines~12--15 are essentially a repetition of lines~8--11, except that the reference image stack is now 
compared with itself. This step is beneficial for estimating a bias in the calculation of the shifts 
\texttt{ux}=$\hat{u}_x$, \texttt{uy}=$\hat{u}_y$ (line~16). This step is explained in more detail in 
subsection~\ref{sec:bias}. 

\section{Derivation of the interpolation kernel $B \star B$}\label{sec:appendix_int_kernel}
It can be shown that all cost function terms can be expressed as sums of cross-correlations of two 
images.

\begin{equation}
	\begin{split}
		C_{DE}(\br; \bu) = \sum_{m=1}^{M} \sum_{w_x=-N}^{N} \sum_{w_y=-N}^{N}
		\underset{=\tilde{D}_m^{(\br)}(\br+\bw)}{\underbrace{\Gamma(\bw) D_m(\br+\bw)}} 
		E_m(\br+\bw+\bu)\\
		= \sum_{m=1}^{M} \sum_{w_x=-\infty}^{\infty} \sum_{w_y=-\infty}^{\infty} 
		\tilde{D}_m^{(\br)}(\br+\bw) E_m(\br+\bw+\bu) = 
		\sum_{m=1}^{M} \left[\tilde{D}_m^{(\br)} \star E_m \right] (\bu).
	\end{split}
\end{equation}

The summation range can be formally extended to infinity because $\Gamma(\bw)$, and thus 
$\tilde{D}_m^{(\br)}(x,y)$, is zero outside of the original summation range.
To reproduce the cost function terms $l_1, \ldots, l_6$, $D_m$ and
$E_m$ are set equal to either $I_{0,m}$, $I_m$, or $\langle I_{0,m} \rangle$.

The following calculation assumes that the function being interpolated is a
summation of such terms. However, this is not the case for the cost functions
already minimised for $T$ [see Eq.~\eqref{eq:LT_short_final}], or $T$ and $D$ 
[Eq.~\eqref{eq:alpha_beta_matrix}, \eqref{eq:TD_alpha_beta} inserted in Eq.~\eqref{eq:LTD}].
Since these equations are non-linear combinations of the cost function terms,
convolution with the found kernel is not strictly equivalent to a cost function
calculation of bi-linearly interpolated images.
Furthermore, the following implies that not $D$ and $E$ are being interpolated,
but $\tilde{D}=\Gamma D$ and $E$. This is not quite equivalent to the ``natural'' way
of calculating cross-correlations of interpolated data, i.e., interpolating $\Gamma$,
$D$, and $E$ separately, and calculating the cost function terms from these.
However, we have not been able to find an equivalent expression of this that
uses only a single convolution.

$C_{DE}$ is only defined for integer values of $u_x$ and $u_y$, but a continuous
form of this which is defined for non-integer shifts can be derived by bilinear
interpolation.
If we assume $D$ and $E$ to be defined on $\mathbb{R}^2$, e.g. as a grid of Dirac
impulses:
\begin{equation}
	\begin{split}
		D(\br) = \sum_{i,j} D_{ij} \delta(\br - [i, j]^T),
	\end{split}
\end{equation}
we can express bilinear interpolation as a convolution of $D$ with the
bilinear interpolation kernel $B$:
\begin{equation}
	\begin{split}
		\widehat{D}(\br) = (D \otimes B)(\br) = \iint d^2\br' D(\br') B(\br - \br').
	\end{split}
\end{equation}
By extension,  the cross-correlation function of the two interpolated functions
$\widehat{D}$, $\widehat{E}$ then is
\begin{equation}
	\begin{split}
		(\widehat{D} \star \widehat{E})(\br) = [(D \otimes B) \star (E \otimes B)](\br) =
		(D \otimes B)(-\br) \otimes (E \otimes B)(\br) = \\
		D(-\br) \otimes B(-\br) \otimes E(\br) \otimes B(\br) =
		D(-\br) \otimes E(\br) \otimes B(-\br) \otimes B(\br) = \\
		[D \star E](\br) \otimes [B \star B](\br).
	\end{split}
\end{equation}

\begin{figure}[ht]
	\centering\includegraphics{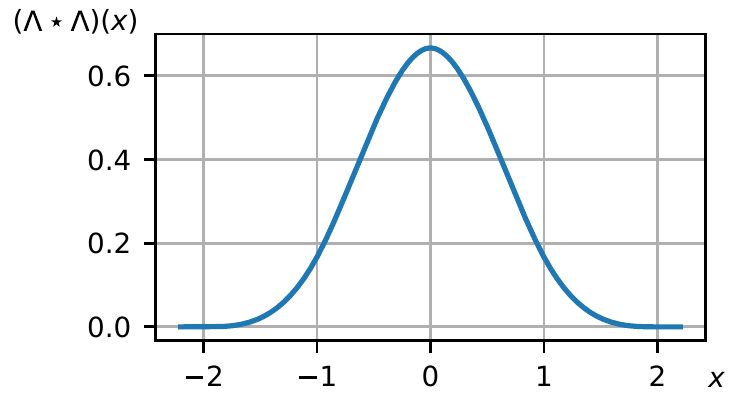}
	\caption{Curve shape of $(\Lambda \star \Lambda)(x)$, the autocorrelation function of the linear 
		interpolation kernel $\Lambda(x)$. For the calculation of the continuous ``cost landscape'', the 
		grid of cost function values for discrete-valued shifts $\bu$ is convolved with the function ${(B 
			\star B)(\br) = (\Lambda \star \Lambda)(r_x) \cdot (\Lambda \star \Lambda)(r_y)}$.}
	\label{fig:bb_kernel}
\end{figure}

\sloppy The above calculation uses the fact that ${(f \star g)(\br) = f(-\br) \otimes g(\br)}$
(for real-valued $f$), that ${(f \otimes g)(-\br) = f(-\br) \otimes g(-\br)}$, and that
${f \otimes g = g \otimes f}$.
In short, the cross-correlation of the bilinear-interpolated versions of $D$ and
$E$ can be derived from their discrete-domain cross-correlation by convolution
with the kernel $B\star B$ (i.e., the autocorrelation of $B$).
Since 
\begin{equation}
	\begin{split}
		\label{eq:def_bilinear}
		B(\br)=\Lambda(r_x) \Lambda(r_y),
		\Lambda(x) = \begin{cases}
			1+x, &-1 \leq x \leq 0, \\
			1-x, &0 \leq x \leq 1, \\
			0 \ \text{else}.
		\end{cases}
	\end{split}
\end{equation}
it follows that
\begin{equation}
	\begin{split}
		\label{eq:full_kernel}
		&(B \star B)(\br) = (\Lambda \star \Lambda)(r_x) \cdot (\Lambda \star \Lambda)(r_y).
	\end{split}
\end{equation}
It is thus sufficient to solve the one-dimensional cross-correlation problem:
\begin{equation}
	\begin{split}
		\label{eq:lambda_cc}
		(\Lambda \star \Lambda)(x) = \int_{-\infty}^{\infty} dx' \Lambda(x') \Lambda(x'+x).
	\end{split}
\end{equation}
It is evident from Eq.~\eqref{eq:def_bilinear} that $(\Lambda \star \Lambda)(x)=0$ for $|x| > 2$, 
and that, since $\Lambda$ is symmetric, $\Lambda \star \Lambda$ is symmetric as well. For the 
remaining cases, Eq.~\eqref{eq:lambda_cc} can be solved by splitting it into intervals according to 
the cases in Eq.~\eqref{eq:def_bilinear}.
For $0 \leq x \leq 1$, these intervals are $[x-1,0]$, $[0,x]$, and $[x,1]$:
\begin{equation}
	\begin{split}
		(\Lambda \star \Lambda)(x) &= \int_{x-1}^{0} (1+x')(1+x'-x)  dx' +
		\int_{0}^{x} (1-x')(1+x'-x) dx' \\
		&+ \int_{x}^{1} (1-x')(1-x'+x) dx' \\
		&= \frac{1}{6}\left( 3 x^3 - 6 x^2 + 4 \right).
	\end{split}
\end{equation}
For $1 \leq x \leq 2$, the only nonzero part of the integral is
\begin{equation}
	\begin{split}
		(\Lambda \star \Lambda)(x) = \int_{x-1}^{1} (1-x')(1+x'-x) = \frac{1}{6} \left( -x^3 + 6 x^2 
		-12 x + 8 \right)
	\end{split}
\end{equation}
The cases $-2 \leq x \leq -1$ and $-1 \leq x \leq 0$ are easily derived using the symmetry of the 
problem, i.e. knowing that $(\Lambda \star \Lambda)(-x) = (\Lambda \star \Lambda)(x)$, we can 
substitute $x \rightarrow -x$, yielding the final result of
\begin{equation}
	\begin{split}
		(\Lambda \star \Lambda)(x) =
		\begin{cases}
			\frac{1}{6} ( x+2 )^3, &-2 \leq x \leq -1, \\
			\frac{1}{6} ( -3 x^3 - 6 x^2 + 4 ), &-1 \leq x \leq 0, \\
			\frac{1}{6} ( 3 x^3 - 6 x^2 + 4 ), &0 \leq x \leq 1, \\
			-\frac{1}{6} ( x-2 )^3, &1 \leq x \leq 2, \\
			0 \ \text{else}.
		\end{cases}
	\end{split}
\end{equation}
The shape of this curve is illustrated in Fig.~\ref{fig:bb_kernel}.

\FloatBarrier

\begin{backmatter}
\bmsection{Funding}
This publication is part of a project that has received funding from the European Research Council (ERC) under the European Union’s Horizon 2020 research and innovation program (Grant agreement No.~866026).

\bmsection{Acknowledgments}
We thank Dr.~Irene Zanette for helpful discussions.
We would like to thank Professor Julia Herzen and Mirko Riedel for the invitation to their beamtime at P05/PETRA~III in October~2021, which allowed us to perform the measurement shown in Fig.~3.
We acknowledge DESY (Hamburg, Germany), a member of the Helmholtz Association HGF, for the provision of experimental facilities. Parts of this research were carried out at PETRA~III and we would like to thank Mirko Riedel and Dr.~Felix Beckmann for assistance in using P05. Beamtime was allocated for proposal II-20190765.
We acknowledge Elettra Sincrotrone Trieste for providing access to its synchrotron radiation facilities and we thank Dr.~Giuliana Tromba and Dr.~Adriano Contillo for assistance in using the SYRMEP beamline. Beamtime was allocated for proposal~20210351.

\bmsection{Disclosures}
The authors declare no conflicts of interest.

\bmsection{Data Availability}
The code for the presented software package is available 
at:\\\url{https://github.com/optimato/UMPA} and at Ref.~\cite{umpa_zenodo}.

\end{backmatter}


\bibliography{umpa_paper}

\begin{thebibliography}{10}
\newcommand{\enquote}[1]{``#1''}

\bibitem{Momose2005}
A.~Momose, \enquote{{Recent Advances in X-ray Phase Imaging},}
  {\protect\JournalTitle{Japanese Journal of Applied Physics}} \textbf{44},
  6355--6367 (2005).

\bibitem{Snigirev1995}
A.~Snigirev, I.~Snigireva, V.~Kohn, S.~Kuznetsov, and I.~Schelokov, \enquote{On
  the possibilities of x-ray phase contrast microimaging by coherent
  high-energy synchrotron radiation,} {\protect\JournalTitle{Rev. Sci.
  Instrum.}} \textbf{66}, 5486--5492 (1995).

\bibitem{Cloetens1999}
P.~Cloetens, W.~Ludwig, J.~Baruchel, D.~V. Dyck, J.~V. Landuyt, J.~P. Guigay,
  and M.~Schlenker, \enquote{Holotomography: Quantitative phase tomography with
  micrometer resolution using hard synchrotron radiation x rays,}
  {\protect\JournalTitle{Applied Physics Letters}} \textbf{75}, 2912--2914
  (1999).

\bibitem{Paganin2002}
D.~Paganin, S.~C. Mayo, T.~E. Gureyev, P.~R. Miller, and S.~W. Wilkins,
  \enquote{Simultaneous phase and amplitude extraction from a single defocused
  image of a homogeneous object,} {\protect\JournalTitle{Journal of
  Microscopy}} \textbf{206}, 33--40 (2002).

\bibitem{Gureyev2020}
T.~E. Gureyev, D.~M. Paganin, B.~Arhatari, S.~T. Taba, S.~Lewis, P.~C. Brennan,
  and H.~M. Quiney, \enquote{Dark-field signal extraction in propagation-based
  phase-contrast imaging,} {\protect\JournalTitle{Physics in Medicine and
  Biology}} \textbf{65}, 215029 (2020).

\bibitem{Chapman1997}
D.~Chapman, W.~Thomlinson, R.~E. Johnston, D.~Washburn, E.~Pisano, N.~Gmür,
  Z.~Zhong, R.~Menk, F.~Arfelli, and D.~Sayers, \enquote{{Diffraction Enhanced
  X-ray Imaging},} {\protect\JournalTitle{Physics in Medicine and Biology}}
  \textbf{42}, 2015--2025 (1997).

\bibitem{Oltulu2003}
O.~Oltulu, Z.~Zhong, M.~Hasnah, M.~N. Wernick, and D.~Chapman,
  \enquote{Extraction of extinction, refraction and absorption properties in
  diffraction enhanced imaging,} {\protect\JournalTitle{Journal of Physics D:
  Applied Physics}} \textbf{36}, 2152--2156 (2003).

\bibitem{Momose2003}
A.~Momose, S.~Kawamoto, I.~Koyama, Y.~Hamaishi, K.~Takai, and Y.~Suzuki,
  \enquote{{Demonstration of X-Ray Talbot Interferometry},}
  {\protect\JournalTitle{Japanese Journal of Applied Physics}} \textbf{42},
  L866--L868 (2003).

\bibitem{Pfeiffer2006}
F.~Pfeiffer, T.~Weitkamp, O.~Bunk, and C.~David, \enquote{Phase retrieval and
  differential phase-contrast imaging with low-brilliance {X-ray} sources,}
  {\protect\JournalTitle{Nature Physics}} \textbf{2}, 258--261 (2006).

\bibitem{Pfeiffer2008}
F.~Pfeiffer, M.~Bech, O.~Bunk, P.~Kraft, E.~F. Eikenberry, C.~Brönnimann,
  C.~Grünzweig, and C.~David, \enquote{Hard-{X-ray} dark-field imaging using a
  grating interferometer,} {\protect\JournalTitle{Nature Materials}}
  \textbf{7}, 134--137 (2008).

\bibitem{Olivo2001}
A.~Olivo, F.~Arfelli, G.~Cantatore, R.~Longo, R.~H. Menk, S.~Pani, M.~Prest,
  P.~Poropat, L.~Rigon, G.~Tromba, E.~Vallazza, and E.~Castelli, \enquote{An
  innovative digital imaging set-up allowing a low-dose approach to phase
  contrast applications in the medical field,} {\protect\JournalTitle{Medical
  Physics}} \textbf{28}, 1610--1619 (2001).

\bibitem{Munro2012}
P.~R. Munro, K.~Ignatyev, R.~D. Speller, and A.~Olivo, \enquote{Phase and
  absorption retrieval using incoherent x-ray sources,}
  {\protect\JournalTitle{Proc. Natl. Acad. Sci. U.S.A.}} \textbf{109},
  13922--13927 (2012).

\bibitem{Endrizzi2014}
M.~Endrizzi and A.~Olivo, \enquote{Absorption, refraction and scattering
  retrieval with an edge-illumination-based imaging setup,}
  {\protect\JournalTitle{Journal of Physics D: Applied Physics}} \textbf{47},
  505102 (2014).

\bibitem{Morgan2010}
K.~S. Morgan, D.~M. Paganin, and K.~K.~W. Siu, \enquote{Quantitative x-ray
  phase-contrast imaging using a single grating of comparable pitch to sample
  feature size,} {\protect\JournalTitle{Optics Letters}} \textbf{36}, 55--57
  (2010).

\bibitem{Gustschin2021}
A.~Gustschin, M.~Riedel, K.~Taphorn, C.~Petrich, W.~Gottwald, W.~Noichl,
  M.~Busse, S.~E. Francis, F.~Beckmann, J.~U. Hammel, J.~Moosmann, P.~Thibault,
  and J.~Herzen, \enquote{High-resolution and sensitivity bi-directional x-ray
  phase contrast imaging using {2D Talbot} array illuminators,}
  {\protect\JournalTitle{Optica}} \textbf{8}, 1588--1595 (2021).

\bibitem{Berujon2012a}
S.~B{\'{e}}rujon, E.~Ziegler, R.~Cerbino, and L.~Peverini,
  \enquote{{Two-Dimensional X-Ray Beam Phase Sensing},}
  {\protect\JournalTitle{Physical Review Letters}} \textbf{108}, 158102 (2012).

\bibitem{Berujon2012b}
S.~Berujon, H.~Wang, and K.~Sawhney, \enquote{X-ray multimodal imaging using a
  random-phase object,} {\protect\JournalTitle{Physical Review A}} \textbf{86},
  063813 (2012).

\bibitem{Morgan2012}
K.~S. Morgan, D.~M. Paganin, and K.~K.~W. Siu, \enquote{X-ray phase imaging
  with a paper analyzer,} {\protect\JournalTitle{Applied Physics Letters}}
  \textbf{100}, 124102 (2012).

\bibitem{Cerbino2008}
R.~Cerbino, L.~Peverini, M.~A.~C. Potenza, A.~Robert, P.~B{\"{o}}secke, and
  M.~Giglio, \enquote{X-ray-scattering information obtained from near-field
  speckle,} {\protect\JournalTitle{Nature Physics}} \textbf{4}, 238--243
  (2008).

\bibitem{Zdora2018}
M.-C. Zdora, \enquote{State of the {Art} of {X}-ray {Speckle}-{Based}
  {Phase}-{Contrast} and {Dark}-{Field} {Imaging},}
  {\protect\JournalTitle{Journal of Imaging}} \textbf{4}, 60 (2018).

\bibitem{Berujon2020b}
S.~Berujon, R.~Cojocaru, P.~Piault, R.~Celestre, T.~Roth, R.~Barrett, and
  E.~Ziegler, \enquote{X-ray optics and beam characterization using random
  modulation: experiments,} {\protect\JournalTitle{Journal of Synchrotron
  Radiation}} \textbf{27}, 293--304 (2020).

\bibitem{Morgan2020b}
A.~J. Morgan, H.~M. Quiney, S.~Bajt, and H.~N. Chapman, \enquote{Ptychographic
  {X-ray} speckle tracking,} {\protect\JournalTitle{Journal of Applied
  Crystallography}} \textbf{53}, 760--780 (2020).

\bibitem{Morgan2020a}
A.~J. Morgan, K.~T. Murray, M.~Prasciolu, H.~Fleckenstein, O.~Yefanov,
  P.~Villanueva-Perez, V.~Mariani, M.~Domaracky, M.~Kuhn, S.~Aplin, I.~Mohacsi,
  M.~Messerschmidt, K.~Stachnik, Y.~Du, A.~Burkhart, A.~Meents, E.~Nazaretski,
  H.~Yan, X.~Huang, Y.~S. Chu, H.~N. Chapman, and S.~Bajt,
  \enquote{{Ptychographic X-ray speckle tracking with multi-layer Laue lens
  systems},} {\protect\JournalTitle{Journal of Applied Crystallography}}
  \textbf{53}, 927--936 (2020).

\bibitem{Paganin2019}
D.~M. Paganin and K.~S. Morgan, \enquote{X-ray {Fokker{\textendash}Planck}
  equation for paraxial imaging,} {\protect\JournalTitle{Sci. Rep.}}
  \textbf{9}, 17537 (2019).

\bibitem{Morgan2019}
K.~S. Morgan and D.~M. Paganin, \enquote{Applying the
  {Fokker{\textendash}Planck} equation to grating-based x-ray phase and
  dark-field imaging,} {\protect\JournalTitle{Sci. Rep.}} \textbf{9}, 17465
  (2019).

\bibitem{Pavlov2020}
K.~M. Pavlov, D.~M. Paganin, H.~T. Li, S.~Berujon, H.~Roug{\'{e}}-Labriet, and
  E.~Brun, \enquote{X-ray multi-modal intrinsic-speckle-tracking,}
  {\protect\JournalTitle{Journal of Optics}} \textbf{22}, 125604 (2020).

\bibitem{Pavlov2021}
K.~M. Pavlov, D.~M. Paganin, K.~S. Morgan, H.~T. Li, S.~Berujon, L.~Quénot,
  and E.~Brun, \enquote{Directional dark-field implicit x-ray speckle tracking
  using an anisotropic-diffusion {Fokker}-{Planck} equation,}
  {\protect\JournalTitle{Physical Review A}} \textbf{104}, 053505 (2021).

\bibitem{Berujon2020a}
S.~Berujon, R.~Cojocaru, P.~Piault, R.~Celestre, T.~Roth, R.~Barrett, and
  E.~Ziegler, \enquote{X-ray optics and beam characterization using random
  modulation: theory,} {\protect\JournalTitle{Journal of Synchrotron
  Radiation}} \textbf{27}, 284--292 (2020).

\bibitem{Pan2009}
B.~Pan, K.~Qian, H.~Xie, and A.~Asundi, \enquote{Two-dimensional digital image
  correlation for in-plane displacement and strain measurement: a review,}
  {\protect\JournalTitle{Measurement Science and Technology}} \textbf{20},
  062001 (2009).

\bibitem{Berujon2015}
S.~Berujon and E.~Ziegler, \enquote{Near-field speckle-scanning-based x-ray
  imaging,} {\protect\JournalTitle{Physical Review A}} \textbf{92}, 013837
  (2015).

\bibitem{Berujon2016}
S.~Berujon and E.~Ziegler, \enquote{X-ray multimodal tomography using
  speckle-vector tracking,} {\protect\JournalTitle{Physical Review Applied}}
  \textbf{5}, 044014 (2016).

\bibitem{Zanette2014}
I.~Zanette, T.~Zhou, A.~Burvall, U.~Lundström, D.~Larsson, M.~Zdora,
  P.~Thibault, F.~Pfeiffer, and H.~Hertz, \enquote{Speckle-based x-ray
  phase-contrast and dark-field imaging with a laboratory source,}
  {\protect\JournalTitle{Physical Review Letters}} \textbf{112}, 253903 (2014).

\bibitem{Zdora2017}
M.-C. Zdora, P.~Thibault, T.~Zhou, F.~J. Koch, J.~Romell, S.~Sala, A.~Last,
  C.~Rau, and I.~Zanette, \enquote{X-ray {Phase}-{Contrast} {Imaging} and
  {Metrology} through {Unified} {Modulated} {Pattern} {Analysis},}
  {\protect\JournalTitle{Physical Review Letters}} \textbf{118}, 203903 (2017).

\bibitem{Behnel2011}
S.~Behnel, R.~Bradshaw, C.~Citro, L.~Dalcin, D.~S. Seljebotn, and K.~Smith,
  \enquote{{Cython: The Best of Both Worlds},} {\protect\JournalTitle{Computing
  in Science and Engineering}} \textbf{13}, 31--39 (2011).

\bibitem{Kottler2007}
C.~Kottler, F.~Pfeiffer, O.~Bunk, C.~Gr\"{u}nzweig, and C.~David,
  \enquote{Grating interferometer based scanning setup for hard x-ray phase
  contrast imaging,} {\protect\JournalTitle{Rev. Sci. Instrum.}} \textbf{78},
  043710 (2007).

\bibitem{Smith2022}
R.~Smith, F.~De~Marco, L.~Broche, M.-C. Zdora, N.~W. Phillips, R.~Boardman, and
  P.~Thibault, \enquote{X-ray directional dark-field imaging using {Unified}
  {Modulated} {Pattern} {Analysis},} {\protect\JournalTitle{PLOS ONE}}
  \textbf{17}, e0273315 (2022).

\bibitem{umpa_zenodo}
F.~De~Marco, P.~Thibault, R.~Smith, and S.~Savatovi\'{c},
  \enquote{{optimato/UMPA: Initial Release of C++ Version},}  (2022). Zenodo,
  \url{https://doi.org/10.5281/zenodo.6984740}.

\end{thebibliography}

\end{document}